%% file: monelli_rev.tex
\documentstyle[12pt,aaspp4]{article}

\newcommand{\msun}{$M/M_{\odot}\,$}

\lefthead{TBD et al.}
\righthead{The Carina Project}

\begin{document}

\title{The Carina Project: II. Stellar Populations\footnote{Based on 
observations collected at the European Southern Observatory, La Silla, 
Chile on Osservatorio Astronomico di Capodimonte guaranteed time.}}

\author{M. Monelli$^{1,2}$, L. Pulone$^1$, C. E. Corsi$^1$, M. Castellani$^1$,
G. Bono$^1$,   A. R. Walker$^3$, E. Brocato$^4$, R. Buonanno$^{1,2}$, 
F. Caputo$^1$, V. Castellani$^{1,5}$, M. Dall'Ora$^{1,2}$, M. Marconi$^6$, 
M. Nonino$^7$, V. Ripepi$^6$, H. A. Smith$^8$}

\affil{1. INAF - Osservatorio Astronomico di Roma, via Frascati 33,
00040 Monte Porzio Catone, Italy; bono, buonanno, caputo, corsi, dallora, 
mkast, monelli, pulone, vittorio@mporzio.astro.it}
\affil{2. Universit\`a  Tor Vergata, Via della Ricerca Scientifica 1, 
00133 Rome, Italy}  
\affil{3. Cerro Tololo Inter-American Observatory, National Optical
Astronomy Observatories\footnote{Operated by Associations of Universities 
for Research in Astronomy (AURA), Inc., under cooperative agreement with 
the National Science Foundation.}, Casilla 603, La Serena, Chile; 
awalker@noao.edu}
\affil{4. INAF - Osservatorio Astronomico di Teramo, via M. Maggini,
64100 Teramo, Italy; brocato@te.astro.it}
\affil{5. INFN - Sezione di Ferrara, via Paradiso 12, 44100 Ferrara, Italy.} 
\affil{6. INAF - Osservatorio Astronomico di Capodimonte, via Moiariello 16,
80131 Napoli, Italy; marcella, ripepi@na.astro.it}
\affil{7. INAF - Osservatorio Astronomico di Trieste, via G.B. Tiepolo 11,
40131 Trieste, Italy; nonino@ts.astro.it}
\affil{8. Dept. of Physics and Astronomy, Michigan State University,
East Lansing, MI 48824, USA; smith@pa.msu.edu}

\date{Received ; accepted  }

\pagebreak
\begin{abstract}
We present a new (V,B-V) Color-Magnitude Diagram (CMD) of the Carina 
dwarf spheroidal (dSph) that extends from the tip of the Red Giant 
Branch (RGB) down to $V\sim25$ mag. Data were collected with the 
Wide Field Imager (WFI) available at 2.2m ESO/MPI telescope and 
cover an area of $\approx0.3$ degree$^2$ around the center of the galaxy.  
We confirm the occurrence of a substantial amount of old stars with
ages around 11 Gyr, together with an intermediate-age population
around 5 Gyr (Smecker-Hane et al. 1994,1996; Hurley-Keller,
Mateo, \& Nemec 1998). Moreover, we also detected a new well-defined blue
plume of young Main-Sequence (MS) stars with an age at most of the order
of 1 Gyr.
This finding is further supported by the detection 
of a sizable sample of Anomalous Cepheids (ACs), whose occurrence can 
be understood in terms of stars with ages $\approx 0.6$ Gyr. The evidence 
for such a young population appears at odds with current cosmological 
models which predict that the most recent star formation (SF) episodes in 
dSph should have taken place 2-3 Gyr ago. 
At odds with previous results available in the literature we found that
stars along the RGB of the old and of the intermediate-age stellar population
indicate a mean metallicity roughly equal to Z=0.0004 ($[Fe/H]\approx -1.7$)
and a small dispersion around this value.
This finding is further strengthened by the reduced spread in luminosity 
of RR Lyrae and Horizontal Branch (HB) stars in the old stellar population 
and of the Red Clump (RC) in the intermediate-age one.

We find evidence for a smooth spatial distribution of the intermediate-age 
stellar population ($\approx 5$ Gyr), which appears more centrally 
concentrated than the oldest one ($\approx 11$ Gyr). The radial 
distribution of the old population appears more clumpy, with a peak  
off-centered by $\approx 2$ arcmin when compared with the Carina 
center. Star counts show a well-defined {\em shoulder} in the 
North-East direction along both the minor and the major axis. 
Current data do not allow us to assess whether this feature is the 
the break in the slope of star count profiles predicted by 
Johnston, Sigurdsson \& Hernquist (1999).    
\end{abstract}


\pagebreak
\section{Introduction}

The dSph galaxies are key stellar systems in several long-standing 
astrophysical problems (Gallagher \& Wyse 1994; Mateo 1998, 
hereinafter M98; van den Bergh 2000), and are fundamental laboratories 
to investigate the evolutionary history of Local Group (LG) galaxies.
According to the evidence of a rather large population of RR Lyrae
variables, these systems were early considered as a sort of nearby 
low-density globular clusters. However, more recent investigations 
have already revealed that several dSphs have experienced a much more 
complex evolutionary history than Galactic Globular Clusters (GGCs) 
do (see e.g. Da Costa 1999). The stellar population content of dwarf 
galaxies has thus become a target of great relevance.

However, the limited area covered by astronomical detectors hampered 
for a long time an exhaustive investigation of these extended celestial 
objects. During the last few years the use of wide field detectors has 
provided an opportunity to sample the bulk of their stellar populations. 
To take advantage of such an opportunity we planned to use the WFI 
available at 2.2m ESO/MPI telescope to investigate stellar 
populations in Carina. The main idea is to supply 
a complete census of both static and variable stars by
collecting time series data covering the pulsation cycle of
evolved variables such as RR Lyrae and Anomalous Cepheids. The
reasons why we selected this dSph are manifold. It is located in
the southern sky, it presents evidence for multiple stellar
populations  (Mould \& Aaronson 1983; Mighell 1990,1997; 
Smecker-Hane et al. 1994,1996; Hurley-Keller et al. 1998; 
Hernandez et al. 2000; Dolphin 2002), and the occurrence of 
RR Lyrae variables has been
already proved (Saha, Monet \& Seitzer 1986, hereinafter SMS). 
Moreover, recent measurements suggest that its core radius is 
roughly 10 arcmin, while the tidal radius is approximately 30 arcmin
(Irwin \& Hatzidimitriou 1995, hereinafter IH). This means that with 
a single field of about $30\times30$ arcmin one collects a substantial 
fraction of its stellar content.

The first deep photometric investigation ($V\sim25$) of Carina dates
back to Mighell (1990). He collected V,R images of an off-center field
($3'.5\times5'.6$) with EFOSC at the ESO 3.6m telescope and found that the
old population could account for approximately the 15\% of the Carina stellar
content. A detailed analysis of the Carina bright stellar component
(V$\approx$22) was provided by Smecker-Hane et al. (1994) who used the
CTIO 1.5m telescope to cover with a mosaic of B,I images the central
$30'.2\times24'.6$ region. These data were supplemented with deep B,R 
data collected by Smecker-Hane et al. (1996) with the CTIO 4m telescope 
and reaching a limiting magnitude of R$\approx$25. The new data disclosed 
for the first time the region across the MS Turn Off (TO) of both an 
intermediate and an old population. This occurrence was confirmed by 
Hurley-Keller et al. (1998) on the basis of deep B,V images
(V$\approx$25) of three off-center fields, covering a sky area of
$\approx 15$ arcmin$^2$ each. Deep V,I data have been collected by 
Mighell (1997) with the Hubble Space Telescope (HST) down to a limiting 
magnitude of $V\sim 27$ mag. Although these data provided an accurate 
CMD well below the TO of the old population the main conclusions of this 
investigation were partially hampered by the limited size of the stellar 
sample, due to the low central stellar density and to the small 
field-of-view covered by the WFPC2 ($2'.4\times2'.4$). The quoted 
investigations
brought forward two morphologically distinct HBs and the occurrence of 
multiple TOs, interpreted by the previous authors as an evidence for 
SF episodes of approximately 2, 3-6  (intermediate-age population), 
and 11-13 (old population) Gyr.
In a recent photometric investigation, Harbeck et al. (2001) using 
evolved Helium burning stars (HB,RC) as stellar tracers found a 
 radial age gradient between old (HB) and intermediate-age (RC) 
stars. This analysis was based on Washington system C,$T_1$ data collected 
with the CTIO 1.5m telescope that cover two rectangular regions of
$14'\times25'$ along the major and the minor axis.
Note that the occurrence of spatial variations in age and in metallicity
in a large sample of LG dwarf galaxies was already suggested by
Grebel (1999) on the basis of HST data.

In this paper we present the results of our photometry which produced
B,V magnitudes for about 68,000 objects, down to V$\sim$25.5.
In \S 2, we discuss the photometric data set collected with the WFI,
together with the reduction procedures and the evaluation of
intrinsic errors (\S 2.1). The approach adopted to transform stellar 
positions from pixels to equatorial coordinates are presented in 
\S 2.2. The estimate of the relative photometric zero-point
of the 8 individual chips is outlined in \S 3, together with the
absolute zero-point calibration. The main features of the CMD are 
discussed in \S 4, while \S 5 deals with the radial distribution of 
selected evolutionary phases. A brief summary of the results and the 
future perspectives of the {\em Carina Project} are outlined in \S 6.

\section{Observations and Data Reduction}

Multiband B and V time series data of the Carina dSph were collected
(E.B., V.C.) for this project with the WFI available at the 2.2m ESO/MPI
telescope (hereinafter WFI) located in La Silla over three consecutive 
nights, from 5th to 7th January 2000. The WFI is a mosaic camera of $2\times4$ 
EEV (English Electric Valve, now e2v technologies. See also 
the web page http://e2vtechnologies.com/) 2046x4098
CCD chips. The pixel scale for this instrument is $0.238^{\prime\prime}$,
therefore each chip covers a sky area of $8'.1\times16'.2$. The observed
field covers an area of $34^{\prime} \times 33^{\prime}$, and was centered
on Carina. Table 1 summarizes positional data together with other
relevant parameters of the Carina dSph. Together with these data we also
adopted a set of B,V images collected for the same project with the Mosaic 
Camera at the CTIO 4m Blanco telescope that cover the
Carina central region as well as a set of B,V,I images collected with
the WFI by the EIS project and available in the ESO Science
Archive\footnote{For more details see http://www.eso.org/science/eis.}
that cover one square degree around the Carina center (see Table 2
for more details). Fig. 1 shows the Carina true color image obtained
by combining the B,V,I median images of the three different data sets.
Current photometric investigation is based on the time series data
collected with the WFI. The CTIO data will be discussed in a forthcoming
paper, while the EIS data have only been adopted to improve the accuracy 
of the astrometric solution (see \S 2.2).

\placefigure{monelli.fig1.ps}

To optimize the sampling of the light curve of radial variables such 
as RR Lyrae and ACs, we secured 54 consecutive B and V exposures of 
$\approx500 s$ each. The pulsation properties of these bright variable 
stars are discussed in a companion paper (Dall'Ora et al. 2003, hereinafter 
Paper I). 
The B,V images were not dithered, because the gaps between the individual
chips of the WFI are relatively small $\sim 23$ arcsec along the Y-axis and
$\sim 7$ arcsec along the X-axis. Moreover, the bright field star HD 48652
(V=9.14) of spectral type F2V was located between the gaps to avoid blooming
effects.
To improve the photometric accuracy of faint stars a few deeper exposures 
were also collected. Table 3\footnote{The complete version of this table 
is only available in the on-line edition of the manuscript.} gives the 
log of the observations together with information on the average seeing 
conditions. One finds that the quality of the images is good, with 
an average seeing $<1.0^{\prime\prime}$ both in the B and in the V band, 
respectively. As a result of both observing strategy and image quality 
we ended up with accurate photometry ($S/N=50$) down to $V\approx 23$ 
and $B\approx 23.5$.

\subsection{Data reduction}

Raw images have been pre-processed by using the NOAO {\it mscred}
(Valdes 1997) tasks available in the IRAF data analysis environment 
for bias subtraction and flat fielding. 
To flat field the data we adopted median sky and dome flats 
collected during the observing nights.
To reject cosmic rays and to improve the detection and measurements 
of faint objects we stacked the eight individual CCD chips using the 
entire set of B and V images, excluding only a few images characterized 
by poor-image quality (seeing $> 1.4^{\prime\prime}$). 
This task was performed by using the following DAOPHOT procedures: 
{\it daomatch}, {\it daomaster}, and {\it montage2} (Stetson 1987, 
1994, private communication). The final eight median images correspond 
to a total exposure time of $27,700 s$ and $28,600 s$ in $V$ and in 
$B$ band, respectively.
Note that the median images are not affected by CCD defects due 
to a shift of a few pixels in the pointing between the individual 
B and V images.

Photometry over the median images was carried out by using the
stand-alone version of the DAOPHOT package (Stetson 1987). The
routine {\it daofind} for star detections was applied to both $V$
and $B$ median images. The detection threshold was fixed at $5
\sigma$ above the local background level. The typical full width
at half maximum (FWHM) of stars in our frames is $\simeq
1^{\prime\prime}$. The list of star candidates was thus cleaned of
extended objects with FWHM larger than $2^{\prime\prime}$ as well
as of spurious identifications. A total of $68,000$ stars were
identified and their fluxes measured with the PSF-fitting
algorithm {\it allstar}. Photometric errors on the mean V and B
magnitudes range from 0.03 for $23 \le V \le 23.5$ to 0.02 for
$23.5 \le B \le 24.0$. This means that current photometry is very
accurate from the tip of the RGB down to the TO of the old stellar 
population.

\subsection{Astrometry}

The transformation of objects from pixel positions to equatorial coordinates 
is a relevant step to properly address the radial distribution of stellar 
populations in Carina and to firmly locate the star candidates for 
forthcoming spectroscopic surveys. We first performed the astrometric 
calibration of a set of eight B-band pointings collected by the EIS project 
that cover 1 square degree around the Carina center (see Tables 2 and 3). 
We adopted the EIS data because they were collected with the WFI, cover 
a wide sky area, and therefore they allowed us to pin point a sizable sample 
of reference stars, and in turn a better accuracy in the astrometric solution.
In particular, we selected $\approx 5000$ reference objects from GSC2 
catalog (McLean et al. 2000) whose magnitude in the $F_{mag}$ band ranges 
from 12 to 19.5. 
The astrometric pipeline we are developing\footnote{Current astrometric 
pipeline (Nonino et al. 1999) is partly based on tools available at 
ftp://ftp.strw.leidenuniv.nl/pub/ldac/software/} was applied to match, 
via the triangulation method (see e.g. Groth 1986), the sources extracted 
from each chip of the EIS images with those present in the GSC2 catalog.

Finally, a third order polynomial 
was used to constrain the full astrometric solution, and this solution was 
used to perform the drizzling of the entire square degree into a 
TAN (tangential) plane.  
Once the astrometric solution for these images has been fixed, 
we used the source extraction package SExtractor (Bertin \& 
Arnouts 1996) to pin point bright but not saturated stars. The 
selected stars were used to anchor the secondary astrometric 
reference system to be adopted for the deeper median B image. 
The astrometric solution for this image 
was obtained using the same approach adopted for the wider ones.
Note that the astrometric solution was now applied to the star 
centroids derived using {\em allstar} (see \S 2.1).  
Fig 2. shows the the spatial distribution of the detected stars in equatorial 
coordinates, clearly disclosing the core radius of the galaxy. The gaps are 
due to the fact that current images were not dithered, while the empty 
circle located close to the center of the galaxy is due to the bright field 
star HD 48652. 

\placefigure{monelli.fig2.ps}

As an independent check of the intrinsic accuracy of current astrometric 
solution, we compared the position of the stars extracted from EIS images 
with those given in the UCAC1 catalog (Zacharias et al. 2000\footnote{See 
also the web page http://ad.usno.navy.mil/ad/ucac/}). By neglecting 
saturated stars, we ended up with a sample of 107 bright stars whose 
coordinates in the UCAC1 catalog were estimated with an accuracy better
than 70 mas. Fig. 3 shows the difference in both right ascension and 
declination between the coordinates in the UCAC1 catalog and those from 
the EIS images routed on the GSC2 catalog. We found that current 
uncertainties on individual 
coordinates appear, at 1 $\sigma$ level, smaller than 100 mas. Note that 
the uncertainties labeled in the figure were estimated over the entire 
sample, i.e. no $\sigma$ clipping has been applied. 

\placefigure{monelli.fig3.ps}

\section{Photometric calibrations}

To accomplish the scientific goal of this project we plan to use
photometric data collected with different CCD cameras and
different telescopes. To improve the accuracy of
the absolute calibration as well as of the relative calibration of
individual chips in wide field imager mosaics, instead of
observing standard stars during each observing night we adopted
the observing strategy of collecting  B and V data for local
secondary stars distributed over our Carina field . Such a
strategy maximizes the use of telescope time and substantially
improves the overall accuracy of different data sets.
Absolute B and V-band photometry of a few dozens of stars located 
across the Carina center was secured during the night of December 
31/2000 (see Tables 4 and 5) by using the 0.9m CTIO telescope. 
However, these local 
secondary standards do not cover the entire field of the WFI, since 
they only cover an area corresponding to the four central CCD chips
of WFI, not the entire field. Moreover, the B-V color 
of these stars is redder than B-V$\sim$0.5, and therefore they do not 
cover the typical colors of blue HB stars and relatively young MS stars.

To improve the accuracy of the relative calibration among the
eight individual CCD chips and to improve the absolute calibration
we performed an independent calibration using data available in
the ESO Science Archive, as provided by two independent sets of 
B,V data of the Carina dSph together with observations of standard 
stars in different Landolt (1992) fields collected with the same 
equipment (WFI, see Tables 4 and 5). 
The former set (December 31/1999 , ESO-064.N-0512) includes 7 standard 
stars in the Landolt field Rubin 149, while the latter one (March 5/2000, 
ESO-064.N-0564) includes standard stars in 4 Landolt fields, namely 
Rubin 149, PG~1633 (5), PG~0918 (4), and PG~1323 (3).

The selection of the above data from the  ESO Science Archive follows a
twofold motivation : {\em i}) B,V data of the same standard field
(Rubin 149) were collected with each of the 8 CCD chips. This means 
that we can properly estimate the response of the individual chips of 
the WFI mosaic. {\em ii}) Previous standard fields do belong to 
the fields for which Stetson (2000\footnote{See also the web page 
http://cadcwww.hia.nrc.ca/standards}) made available to the
astronomical community an extended list of standards. Moreover and
even more importantly, the Carina calibration fields observed
together with these standards are fully (March set) or partially
(December set) overlapped to our scientific data. Table 5 lists
the standard stellar fields adopted in our calibration, the number 
and the time exposures in each band, the name of the standard field, 
as well the pointings on the Carina dSph.

Each data set presents pros and cons; indeed the exposure times of the
first set is of the order of 20-30 sec, standards were observed with each 
of the
8 CCD chips, but only the Rubin~149 field was observed and at very similar
air masses. On the other hand, the second set presents shorter exposure
times (t$\approx$5-10 s), but 4 Landolt fields were observed and at different
air masses. In order to use all the available information and to avoid
deceptive systematic errors in the calibration we decided to perform 
independent calibrations. The reason is because the two calibrations allow
us to cross check both the relative and the absolute zero-points as well as
the color dependence.

\subsection{Relative calibration}

The advent of wide field cameras with CCD mosaics have had a strong impact
on the study of stellar populations in globular clusters and in extragalactic
systems but introduced several calibrations problems that need a careful
investigation well-beyond the classical flat-fielding procedures.
The large field of view coupled with the contemporary use of different CCD
chips and/or readout solutions do not imply that the response of the
detectors is homogeneous within a few percent in different regions of the
field. To constrain relative differences across the WFI camera we
reduced the data of the standard field Rubin 149 collected in Dec 1999, since
it has been observed in each chip with exposure times that allow us to reach
a good S/N ratio down to V$\sim$19 mag.

The observations were secured at almost constant air mass ($\approx 1.15$),
and therefore we can  neglect the extinction corrections as far as the
relative calibration  is concerned. The raw images were pre-processed 
by using the same
procedure adopted for the scientific images (see \S 2.1). The synthetic
aperture photometry was performed with a radius of 30 px ($\simeq$ 7 arcsec).
The analysis of the residuals of the instrumental magnitudes of the same
stars measured in the 8 chips (taking as reference the chip \# 56) shows a
quite odd behavior: the relative difference among the chips shows small
zero-point off-sets of the order of a few hundredths of magnitude but the
scatter was substantially larger than expected according to the random
photometric errors, especially for the four top chips (see Fig. 4).

\placefigure{monelli.fig4.ps}

The residuals did not show any evident correlation with the magnitude
and the color, but we found out that they present a
well-defined dependence on the declination. The effect is of the order
of several hundredths of magnitude at the edges of the WFI and is vanishing
when moving toward the center of the mosaic. A smaller dependence on the RA
might be present but in our opinion it can be neglected because it is at
most of the order of 0.01 mag.
Note that we found the same trend if we adopt the chip \# 51 as a reference,
the only difference is that the scatter now is larger in the 4 bottom chips.
This means that the same star, when observed in different regions of the
WFI can present magnitude differences up to $\sim$ 0.07 mag. Moreover and
even more importantly, we found that this effect changes from chip to chip
and also depends on the photometric band. In particular current data suggest
that the effect is larger in the V than in the B band.
Similar problems were already found by Andersen, Freyhammer, \& Storm 
(1995) and by Manfroid \& Selman (2001). The latter authors used the WFI
and suggested a correction procedure 
called "photometric superflat" that relies on observations of dense 
stellar fields with an {\em ad hoc} dithering of scientific images.

Fortunately, the data of the field Rubin 149 are well distributed over
the 8 chips and allowed us to give an empirical quantitative
estimate of the corrections needed to overcome the problem.
The corrections resulted in a set of polynomials which correct
the instrumental magnitudes of each individual chip as a function
of the Y coordinate. These corrections have been applied to the
standard stars of Rubin 149. Fig. 5 shows the difference between
the V standard magnitudes and the magnitudes measured in the 8 chips,
once the positional effect has been removed, plotted as a function
of the V magnitude. Fig. 6 shows the same residuals but plotted as a
function of the (B-V) color. Data plotted in these figures disclose
quite clearly that the spurious scatter among the individual chips
has been properly removed, and indeed the scatter is homogeneous
and  dominated by the S/N ratio of individual measurements.
Moreover, it is noteworthy that data plotted in Fig. 6 do not
display a significative trend with the (B-V) color.

\placefigure{monelli.fig5.ps}
\placefigure{monelli.fig6.ps}

Figures 7 and 8\footnote{These figures are only available in the on-line 
edition of the manuscript.} show the same data of Figs. 5 and 6, but 
for the the B magnitude.
\placefigure{monelli.fig7.ps}
\placefigure{monelli.fig8.ps}
Data plotted in these figures support the evidence that both the $\Delta V$
and the $\Delta B$ present a trend neither with the magnitude nor with the
B-V color. This suggests that current procedure supplies homogeneous
instrumental magnitudes, within a few percent, wherever the stars are
located inside the WFI mosaic. To investigate whether the previous
corrections change with time we applied the same corrections to the
standard fields observed in March 2000, as well as to the WFI data
for 47~Tuc collected in November 1999 and available at the ESO Science 
Archive.
Interestingly enough, we found that the empirical corrections supply a
good calibration also for these data sets. This suggests that within
this time interval the relative response of the camera did not change.
A more detailed discussion concerning the empirical corrections for the 
different photometric bands will be addressed in a forthcoming paper.

\subsection{Absolute calibration}

To improve the color range covered by standard stars and to estimate
the extinction coefficients we also adopted the set of observations
collected in March 2000 (see Table 5\footnote{This table is only available 
in the on-line edition of the manuscript.}).  
To estimate the extinction coefficients we selected $22 + 11$ 
standard stars brighter than V=17 mag from the Stetson's standards 
(Stetson 2000) located in the
Landolt fields PG1633 and PG0918, since they have been collected at
substantially different air masses. We find that the first order
extinction coefficient are: $k_v=0.13\pm0.02$ and  $k_b=0.25\pm0.02$.
Interestingly enough, these values are quite similar to the mean
values measured at La Silla over the same period of the year (see the
ESO web page http://www.ls.eso.org/lasilla/atm-ext/ and the Geneva 
Observatory web page http://obswww.unige.ch/photom/extlast.html).

As a final step we adopted most of the Stetson's standard stars in the
four Landolt fields brighter than $V=19$. We ended up with a sample
of 117 standard stars in the color range $-0.2 \le B-V \le 1.4$.
The aperture photometry was performed using the same radius adopted
for Rubin~149. We find that the transformations into the Johnson B,V
standard system are:

$V=v{^0_1s}-0.117(\pm0.013)(b{^0_1s}-v{^0_1s})+24.068(\pm0.025)$

$B=b{^0_1s}+0.456(\pm0.016)(b{^0_1s}-v{^0_1s})+24.788(\pm0.033)$

where $v{^0_1s}$ and $b{^0_1s}$ are the instrumental magnitudes scaled
to the same exposure time (1s) and corrected for the extinction as well
as for the positional effect discussed in the previous subsection.
The errors in parentheses are the errors on the coefficients.

Fig. 9 shows the comparison between the Stetson's magnitudes and colors
with those based on current photometry and calibrations.
The comparison does not show any systematic trend with magnitudes
and colors and the agreement is satisfactory. Current uncertainties
are consistent with the S/N ratio of photometric data involved in the
measurements.

\placefigure{monelli.fig9.ps}

The scientific B,V frames centered on Carina collected together with
the Landolt standard fields during the observing run of March 2000
(see Table 3) have been used to provide a set of secondary Johnson
standards to be used for the absolute calibration of the data.
Approximately 60 bright and isolated stars were selected in each
CCD chip, and then measured with aperture photometry, corrected
for the individual positional effect of the chip, and
eventually calibrated with the previous transformations. The Carina
field observed together with the standard stars in December 1999
(see Table 3) partially overlaps with the Carina field observed
in March. The chips \# 50 and \# 57 (March) overlap with the chips
\# 53 and \# 54 (December). The photometry and the absolute calibration
of the December data set were performed by following the same procedure
adopted for the March data set. Fig. 10 displays the difference in
color (top) and in V magnitude (bottom) between the secondary standards
in common between the two data sets. Data plotted in this figure show,
within photometric uncertainties, a good agreement for
$15 \le V \le 19$ mag. This supports the internal consistency of the
absolute calibration.

\placefigure{monelli.fig10.ps}

To calibrate the entire sample of $\sim$68,000 Carina stars distributed
throughout the WFI mosaic we can now follow two different routes:
{\em i)} the individual subsamples of secondary standards once
located in the 8 CCDs can be adopted to supply independent color
equations for each chip; {\em ii)} the instrumental magnitudes of the
secondary standards in the 8 CCD chips are at first homogenized using the
corrections derived in \S 4.1, and then adopted to derive a unique set of
color equations valid for the entire WFI mosaic.
We performed several tests to account for the uncertainties affecting the
two different routes, and eventually we decided to adopt the former one
to avoid the possible errors due to small differences in the PSF adopted
for the various chips.

To check {\em a posteriori} the accuracy of current calibrations we
performed an independent test. We already mentioned that
an independent set of secondary standards are available for this
project that have been collected with different observational equipment.
Fig. 11 shows the difference in  magnitudes and colors for the 25 stars
brighter than V$\sim17.5$ mag in common between the two samples. The data
do not present any magnitude or color dependence, therefore we can conclude
that two fully independent calibrations give the same results within the
errors. This result further strengthens the procedure adopted to transform
the instrumental magnitudes into the Johnson B,V bands.

\placefigure{monelli.fig11.ps}

As a final test we also compared the accuracy of our photometry with the 
B,V data of the Carina galaxy collected by Hurley-Keller et al. (1998). 
The top panel of Fig. 12 shows a large spread in the color
difference but such a difference presents a systematic trend only
for very red stars $(B-V)\ge1.3$. Note that the discrepant points in
the $\Delta (B-V)$ diagram and the discrepant bright points in the
$\Delta V$ diagram are the same objects. The absolute calibration
of the previous data was performed by Mateo et al. (1991) and they
pointed out technical difficulties for the calibration of bright stars.
\placefigure{monelli.fig12.ps}
On the other hand, the difference
in the V magnitude plotted in the bottom panel discloses the
occurrence of a systematic shift roughly equal to 0.05 mag.
The agreement that we found with the independent photometry
discussed in Fig. 11 and the fact that the intrinsic accuracy
of the calibration adopted by Hurley-Keller et al. (1998) is of the
order of $\pm 0.03$ (Mateo et al. 1991) supports the evidence that
current magnitudes are marginally affected by systematic errors.

\section{The Color Magnitude Diagram}

Figure 13 shows the CMD of the eight individual chips. The number of
detected stars traces the position of the galaxy, and indeed the
center is located close to the top left corner of chip n. \#55. In 
the four central chips one finds evidence for the multiple stellar 
populations, and in particular the sequences of H burning stars in
the TO-Sub Giant Branch (SGB) region and of He burning giants, that 
are distributed along an extended and possibly very old HB, as well as 
in a well-defined RC typical of intermediate-mass stars.
\placefigure{monelli.fig13.ps} 
These observational features are further strengthened by data plotted 
in Fig. 14, that shows the CMD for the entire sample of stars. 
\placefigure{monelli.fig14.ps}

On the basis of deep V,I (V$\approx$27) band data collected with HST, 
Mighell (1997, and references therein) suggested that the very old 
population in Carina is a small fraction, if any, of the entire galaxy. 
A similar conclusion was also reached by Hurley-Keller et al. (1998). 
According to these authors the bulk of the stars in Carina formed in 
two different episodes roughly 3 and 7 Gyr ago, with the additional 
evidence of approximately 10-20\% of very old stars with ages 
$\approx 15$ Gyr. However, Hernandez, Gilmore \& Valls-Gabaud (2000) 
on the basis of the same HST data concluded that Carina experienced 
three different SF episodes, namely at 3, 5, and 8 Gyr ago, with no 
evidence of stars older than 10 Gyr. This result has been recently 
questioned by Dolphin (2002) who marginally detected (1 $\sigma$ level) 
older stars. These empirical results support the evidence that to 
properly estimate the fraction of different stellar components in 
dSph galaxies a substantial portion of the body of these systems 
need to be covered.

To figure out the Carina evolutionary properties according to the 
observed CM diagram, one has first to single out detailed information 
on its metal content. This has been done by comparing the distribution 
in the CMD of the oldest Carina population with similar diagrams of old 
stellar systems with known metal abundance. This approach relies on 
the well-known evidence that the slope of the RGB is a bona fide 
metallicity indicator (Hartwick 1968). To our great surprise we found 
that the Carina old stellar population nicely follows the stellar 
distribution of the GGC NGC~1904 (Piotto et al. 2002), for which a 
mean metallicity of [Fe/H]$\approx-1.69$ (Z=0.0004) has been 
estimated (Zinn \& West 1984). 
This empirical evidence strongly suggests not only a similar metallicity, 
but also a similar age, thus constraining the age of Carina old population 
in the range of ages typical of GGCs, i.e. 10-13 Gyr (Vandenberg et al. 2002).
We notice that this metallicity estimate appears in good agreement with 
recent low-resolution spectra of the near infrared Ca II triplet in 
52 stars located close to the tip of the Carina RGB. These data support 
a mean metallicity of $[Fe/H]=-1.99\pm0.08$ (Smecker-Hane et al. 1999). 
However, these measurements present 
an intrinsic spread in metallicity of $\approx 0.25$ dex (1$\sigma$)
and a full width of approximately 1 dex. This suggests that Carina
could underwent a complex chemical evolution and/or SF history 
during the last few Gyrs.
Before we can dig into speculations concerning the occurrence of the
so-called age-metallicity degeneracy (e.g., Hodge 1989; Da Costa 1991)
we need accurate, high resolution spectroscopic measurements of heavy 
element abundances.
Recent findings brought forward systematic differences not only between
photometric indices and spectroscopic observations (Bonifacio et al. 2000;
Cole, Smecker-Hane, \& Gallagher 2000) but also between metallicities based
on Ca II triplet and on Fe-peak element abundances (Tolstoy et al. 2001).
Note that the latter is a thorny problem in stellar systems such as dwarf
galaxies, because Ca is an $\alpha$-element.

An even more compelling comparison can be performed with the old LMC 
cluster Reticulum (Monelli et al. 2002), for which spectroscopic 
measurements support a metal 
abundance of $[Fe/H]=-1.71\pm0.09$  (Z=0.0004, Suntzeff et al. 1992). 
This cluster indeed presents several advantages when compared with 
NGC~1904, the absolute calibration is more accurate, the relevant 
evolutionary phases are very well-defined across the CMD, and in 
particular the HB luminosity level. The comparison of HBs is facilitated 
because the Reticulum HB completely covers the instability strip 
(32 RR Lyrae, Walker 1992), whereas NGC~1904 only contains blue HB stars 
and a few RR Lyrae (4 bona fide and 4 candidates, 
see Ferraro et al. 1992, and references therein). 
Fig. 15 shows that by simply decreasing the apparent magnitude of Reticulum 
stars by $\Delta V\approx 1.68$ mag the old Carina population nicely 
overlaps the Reticulum stellar loci all over H and He burning phase.
One can derive the conclusion that not only the two populations have 
quite similar evolutionary properties, but also that the two stellar 
systems should have quite similar reddenings.
\placefigure{monelli.fig15.ps}
According to the pulsation properties of RR Lyrae stars, Walker (1992) 
estimated a mean reddening for Reticulum of $E(B-V)=0.03\pm0.02$.  
The reddening map provided by Burstein \& Heiles (1982) supplies 
for Carina a mean reddening of 
$E(B-V)\approx 0.025$, while the dust infrared map by Schlegel, 
Finkbeiner, \& Davis (1998) provides, across the central 0.3 degree$^2$ 
covered by our data, a mean reddening value of $E(B-V)=0.058\pm0.013$. 
According to this map the reddening along the Northern direction is
constant, whereas when moving from East to West presents a mild increase.
In the following we assume E(B-V)=0.03, with an uncertainty which should
not be larger than $\pm0.01-0.02$ mag.

Once we fixed both reddening value and metal content, we can compare 
empirical data  with suitable theoretical isochrones. The comparison was 
performed using stellar isochrones from the Pisa Evolutionary Library 
(Castellani et al. 2003, in preparation). The mixing length value adopted 
in these evolutionary models was calibrated as a function of metal content 
on the observed RGBs of GGCs. Theoretical predictions  
were transformed into the observational plane by adopting the bolometric 
corrections and the color-temperature relations predicted by atmosphere 
models (Castelli, Gratton, \& Kurucz 1997a,b). 
To validate physical assumptions adopted to construct current stellar 
isochrones, Fig. 16 shows the fit of the 11 Gyr isochrone (solid line) and 
of the Zero-Age-Horizontal-Branch (ZAHB, dashed line) to empirical data for 
the two quoted globular clusters. 
Theoretical predictions for NGC~1904 were plotted by adopting an apparent 
distance modulus of $DM_V=(m-M)_V$=15.72 mag and a reddening of E(B-V)=0.03 
(Kravtsov et al. 1997). The distance was estimated using the apparent 
magnitude of HB stars estimated by Kravtsov et al. (1997, $V_{HB}=16.25$) 
and the calibration of the $M_V$ vs [Fe/H] relation provided by 
Caputo et al. (2000). The distance modulus we adopted for Reticulum is 
based on the classical Cepheid distance scale (Bono et al. 2002). 
Data plotted in this figure show that theory accounts quite well for
both H and He burning evolutionary phases.  
\placefigure{monelli.fig16.ps}
Now we adopt the same theoretical framework to constrain evolutionary 
properties of Carina stellar populations. 
Fig. 17 shows that theoretical predictions, by assuming a reddening of
E(B-V)=0.03, a mean metallicity of Z=0.0004, and a distance $DM_V$=20.24
nicely fit
the observed distributions for selected assumptions about stellar ages 
(see labeled values). The adopted distance appears in reasonable 
agreement with previous evaluations by Mighell (1997), Mateo, 
Hurley-Keller, \& Nemec (1998), Girardi \& Salaris (2001), and 
by Paper I. As a whole, we found evidence for a substantial amount of 
old stars with ages ranging around $11\pm1$ Gyr. This finding, as well as 
the sizable sample of RR Lyrae stars (75) we have detected, confirms
the early results by SMS and by Smecker-Hane et al. (1994). 

\placefigure{monelli.fig17.ps}

At the same time, data plotted in Fig. 17 show that the bulk of the 
intermediate-mass stars present TO ages in the range of $5\pm2$ Gyr. 
Age estimates for RC stars provided by Caputo, Castellani \&
Degl'Innocenti (1995) on the basis of the difference in luminosity 
with the SGB give quite similar ages. This means that we can safely 
conclude that RC stars in Carina are the bona fide counterpart of the 
above intermediate-age population.

Moreover and even more importantly, data in Fig. 17 bring forward the 
occurrence of young MS stars with ages younger than 1 Gyr 
(see the 0.6 Gyr isochrone  plotted in the figure). The occurrence of 
relatively young stars is interesting because the available literature 
suggested that the most recent SF episode 
in Carina took place approximately 2-3 Gyr ago (Dolphin 2002). In this 
context it is worth mentioning that a stellar population as young as 
this might account for the sizable sample of Anomalous Cepheids 
(6 bona fide plus 9 candidates) we detected (see Paper I).  
If we assume that these bright variables are the progeny of young 
single stars then young, intermediate-mass He burning structures 
should account for their distribution in the CMD. 
To further constrain the nature of these objects we decided to investigate,
as suggested by the referee, whether they could be Blue Stragglers (BSs) of
the old stellar population. Following Fusi Pecci et al. (1992) and Mateo,
Fischer, \& Krzeminski (1995) we estimated the specific frequency of blue
plume stars and we found that $N_B/N_{HB}=360/280\approx1.3$. Recent
empirical estimates suggest that GGCs characterized by very low central
densities present a specific frequency of BSs that is at least a factor of
2.5 smaller, namely $N_{BS}/N_{HB}\approx0.5$ (Preston \& Sneden 2000).
This finding further strengthens the hypothesis that blue plume stars 
are genuine young stars.

Figure 18 shows the comparison between old, low-mass RR Lyrae stars 
(circles) and young, intermediate-mass ACs (triangles) with theoretical 
predictions for Zero Age He-burning structures at fixed chemical 
composition and different assumptions for the progenitor age 
(Cassisi et al. 2003, in preparation).  
Data in this figure support the evidence that the position of bright 
and bluer ACs appear in reasonable agreement with stars originating
from the Zero Age He-burning structures constructed assuming the same 
chemical composition of the old stellar component (Y=0.23, Z=0.0004), 
a progenitor age of 0.6 Gyr, and therefore with a stellar mass 
$M\approx 2.2 M_\odot$. This finding is in good agreement with the 
lower limit in stellar mass provided in Paper I on the basis 
of their pulsation properties, i.e. $M\ge1.4-1.5M_\odot$. However, we 
cannot exclude that some redder ACs are the result of mass transfer in 
old binary systems (Renzini, Mengel, \& Sweigart 1977;  Corwin, Carney, 
\& Nifong 1999). 

\placefigure{monelli.fig18.ps}

To constrain the Carina mean metallicity we performed the same comparison 
between theory and observations of Figures 17 and 18 but using isochrones 
and He burning models for Z=0.0006 ([Fe/H]$\approx-1.5$) and 
Z=0.0002 ([Fe/H]$\approx-2.0$). Interestingly enough, we found that 
the new predictions for plausible assumptions concerning the distance 
and the reddening do not account for the three observables that are 
more sensitive to metal abundance, namely RGB color, HB luminosity, 
and RC color. In particular, more metal-rich predictions account for 
RGB color, and HB luminosity but Zero Age He-burning models are 
systematically bluer than RC stars. On the other hand, the more 
metal-poor predictions account for HB luminosity and RC color, 
but the isochrones for the same ages are systematically redder 
than RGB stars.  

Before concluding this section, let us notice that an age of the order of
11 Gyr makes the Carina old population coeval not only with NGC~1904 and
Reticulum, but also with several other well studied globulars in the Galaxy
(Cassisi et al. 1999) as well as in the Magellanic Clouds (Brocato et al. 
1996). As long as the evidence that these stellar systems are coeval 
is not affected at all by the adopted theoretical evolutionary scenario, 
the evaluation of the absolute age does depend on it. By adopting the same 
theoretical framework but neglecting the efficiency of element diffusion 
we would derive cluster ages older by approximately 1 Gyr.

\section{Radial distributions}

On the basis of preliminary empirical evidence (Mighell 1997; 
Harbeck et al. 2001) it has been suggested that the old and the
intermediate-age populations in Carina present different radial 
distributions. To investigate the spatial distribution of
different stellar populations, we selected stars in suitable boxes
representative of the old (see dashed boxes in Fig. 19) and of the
intermediate-age population (solid boxes). This approach, when compared 
with similar analyses in the literature, presents the substantial
advantage to use quite large stellar samples both for the old
($\approx 1000$) and the intermediate-age ($\approx 5000$) 
populations.
\placefigure{monelli.fig19.ps}
Fig. 20 shows the comparison between the isodensities of the whole
stellar sample (dashed contours) in Carina with those for the old
(HB, SGB) stellar population (solid contours). The isodensity
contours range from $\approx 20\%$, to $100\%$ of the central
maximum density with a density step of 20\%. The central maximum 
density in the central bin ($2\times2$ arcmin squared) is $\approx 1400$ 
for the entire galaxy and $\approx 10$ for old HB stars.
Fig. 21 shows the same data, but the solid contours refer to the young 
(RC, intermediate-age MS) stellar population. The central maximum 
density is $\approx 20$ for RC stars.
\placefigure{monelli.fig20.ps}
\placefigure{monelli.fig21.ps}
Data plotted in these figures show that the isodensity contours of
the intermediate-age population resembles quite well the density
distribution of the whole data. On the contrary, the isodensity
contours of the old population appear more irregular, with a small
offset of about 2 arcmin in the peak density of the old population
when compared with the Carina center (see the cross in Fig. 20 and
21). To estimate on a quantitative basis the difference between
the two populations we performed several two-tails Kolmogorov-Smirnov 
(KS) tests. At first, we tested whether the two old
subsamples, i.e. the HB and SGB, present the same radial
distributions. We found that the two distributions are almost
identical, since the KS probability is equal to 90\%. The same
outcome applies to the two young subsamples, namely RC and
intermediate-age MS, and indeed the KS probability is equal to 94\%.
On the the other hand, the KS test applied to the old and the
young samples, supplies a vanishing probability that the two
radial distributions are the same.

To avoid deceptive errors introduced by the spatial smoothing of the
isocontour levels we decided to investigate the {\em in situ} spatial
distributions of the two populations along the major and the minor
axis, assuming a position angle of approximately 60 degrees ($pa=65\pm5$, IH).
Moving along the major axis and by assuming a bin size of 2 arcmin  
we computed the histogram of the entire stellar content of the galaxy. The
same procedure was repeated but along the minor axis. Fig. 22 shows
the spatial distribution along the major (top) and the minor (bottom) axis
respectively. Interestingly enough, we found that the eccentricity
($e=1-b/a$), where $b$ and $a$ are the FWHMs of the individual distributions
given by the gaussian fits, is roughly equal to $\approx$0.3. This estimate
is in good agreement with the estimate provided by IH.

The same procedure has been applied to the old and the young samples.
Figures 23 and 24 show the two spatial distributions along the major axis
and the minor axis respectively. 
\placefigure{monelli.fig23.ps}
\placefigure{monelli.fig24.ps}
In both cases the spatial distribution of the old component is broader
than the distribution of the young component. It turns out that the
eccentricity of the old component is $\approx0.27$, and therefore quite
similar to the eccentricity of the entire galaxy. On the other hand, the
eccentricity of the young population is $\approx 0.38$.
We performed several tests by changing the bin size from 1.5 to 3 arcmin  
and we found that the eccentricities of the young and old populations 
change by less than a few percent.  
Therefore, current estimates seem to suggest that the old stellar component 
is distributed over a sort of "spheroidal halo", while the young component 
is more concentrated and flattened along the major axis. Moreover, data 
plotted in Fig. 23 also show an interesting feature: the spatial distribution
along the major axis of the old population peaks at $\approx 2\pm1$ arcmin 
from the Carina center, while it is centrally peaked along the minor axis. 
On the contrary, the young population is centrally peaked both along
the major and the minor axis. This evidence further supports the hypothesis  
of a difference in the spatial distribution between the old and the young 
stellar populations.

We mention that data plotted in Fig. 22, also suggest that the spatial 
distribution of the old component along the two axes is not centrally 
symmetric. The slope of the star counts in the South-West direction is 
steeper than along the opposite direction. These empirical evidence seem 
to suggest that the young and more massive population is more centrally
concentrated than the old one.  The present result, once confirmed, 
might provide valuable constraints on the Carina dynamical history 
and on its interaction with the Galaxy (Kuhn, Smith, \& Hawley 1996; 
Majewski et al. 2000).

In a recent photometric investigation Stetson, Hesser, \& Smacker-Hane
(1998) found that different stellar components in the Fornax dSph might
present different centroids. Therefore, we performed a new test to figure 
out whether the young stellar component in Carina exhibits the same 
radial distribution as the intermediate-age one. The KS test suggests that 
blue plume stars present a radial distribution quite similar to RC and 
MS stars, and indeed the KS probability is roughly equal to $\approx 50$\%. 
On the other hand, the KS test with old HB stars indicates that the two 
populations are different, since the KS probability is vanishingly 
(P=0.00001) small. 

\placefigure{monelli.fig25.ps}

Finally, we also investigated the radial distribution of blue plume stars 
along the major and the minor axis. The comparison between data plotted in
Fig. 25 and data plotted in the bottom panel of Fig. 23 and 24 indicates 
that the radial distributions of these two stellar components are quite
similar along the minor axis. Nevertheless there is a mild evidence that
blue plume stars along the major axis appear more centrally concentrated.
Unfortunately, current data do not allow us to constrain on a quantitative 
basis the difference in the centroids.

\section{Summary and final remarks}

This is the second paper of a large program devoted to the evolutionary
and pulsation properties of Carina stellar populations. We
presented B,V time series data collected over three consecutive
nights with the WFI. Special care has been
given to the relative and absolute photometric calibration of the
data. We found that the response of the eight individual chips is
not uniform but depends on the position of the star across the
chip. On the basis of a sizable sample of standard stars observed
with the eight chips we removed the positional effect. To
firmly constrain the photometric zero-point we performed two
independent absolute calibrations using data collected with
different telescopes. The comparison between the two sets of
secondary standards shows a good agreement within the errors.
We investigated the main evolutionary phases of the CMD as a
function of the position and we found that when moving along the
minor axis the radial distribution of young and intermediate-age 
stars decreases more rapidly than the distribution of the old stellar
population. We found that theoretical predictions for Z=0.0004 are
in good agreement not only for MS stars but also for old, low-mass
HB stars as well as for young and intermediate-age He-burning stars.

The comparison between theory and observations brought forward
several interesting findings: \\ 
{\em i}) A well-defined plume of blue
and bright MS stars located at $V\approx22$ and $(B-V)\approx0$
suggest that the last SF episode in Carina took place
not more than 1 Gyr ago. This seems a robust identification, since
this region of the CMD is marginally affected by field
contamination. The occurrence of a young stellar population is
supported by the large sample of ACs we have identified.
If these interesting objects are the evolutionary aftermath of
young single stars, then their distribution in the CMD can be 
explained, according to current theoretical predictions, as central
He-burning models with stellar masses ranging from $\approx 1.4$
to $\approx 2.0$ \msun and stellar ages of the order of 0.6 Gyr.
This evidence together with previous findings on the Fornax dSph
(Stetson et al. 1998; Gallart et al. 2002) are at odds with the dynamical
mechanism, "tidal stirring", suggested by Mayer et al. (2001a,b) to 
account for the dSphs in the Local Group. The numerical
simulations they performed suggest that the SF after 2-3
orbits (approximately 10 Gyr) should be vanishing, since the gas
was either transformed into stars or stripped by tidal shocks.\\ 
{\em ii)} The CMD for magnitudes ranging from V$\approx 22$ to
V$\approx 23$ shows a broad MS and a well-populated SGB. These
features support the evidence that Carina experienced an important
SF episode $6\pm1$ Gyr ago. After this event the SF was less efficient 
but an ongoing process for at least 3-4 Gyr. This evidence is in contrast 
with standard Cold Dark Matter (CDM) theories. In fact, dSphs within this 
theoretical framework were the first galaxies to be formed. Therefore dSphs 
with low-velocity
dispersions ($\le 10$ km/s) should not present any SF activity during
the last 12 Gyr (Moore et al. 1999; Cen 2001). On the other hand,
empirical data strongly support the evidence that a large fraction of
dSphs in the LG have low-velocity dispersions and experienced SF
activity during the last few Gyrs (Caputo et al. 1999;
Piersimoni et al. 1999; Tolstoy et al. 2001, and references therein).\\ 
{\em iii}) The detection of a sizable sample of blue and red HB
stars as well as of RR Lyrae stars confirms the previous findings
by Smecker-Hane et al. (1994) and by Saha, Monet \& Seitzer (1986) 
concerning the occurrence of a old stellar population in Carina. 
Note that the results concerning the occurrence of a young stellar
component in Carina is a consequence of
the fact that current data cover a substantial portion of the body of 
the galaxy, while data sets available in the literature either reach 
brighter limit magnitudes or cover a smaller field of view. In particular, 
the small field of view covered by WFPC2 hampers HST observations due to 
the low stellar density in the central regions of dSphs.\\  
{\em iv}) Magnitude distribution of HB stars as well as of RR Lyrae
stars show, within current uncertainty, a very small spread in
luminosity. This finding together with the luminosity and the color
of RC stars suggests that the spread in metallicity among old and
intermediate-age populations is relatively small.

Thanks to the large field of view covered by our data we also
investigated the radial distribution of old and intermediate-age   
stellar populations. To avoid deceptive errors due to the size of
the sample we selected HB and SGB stars for the old, low-mass
population and RC and MS stars for the intermediate-age stellar
component. Interestingly enough we found that:\\ 
{\em i}) The isodensity contours of the intermediate-age population
properly fit the spatial distribution of the entire galaxy, whereas
the isodensity contours of the old population is somehow more clumpy.\\ 
{\em ii)} The peak of the old stellar population is roughly
2 arcmin off-center when compared with the peak of the intermediate-age  
population. This finding is supported by the spatial distribution
of the two stellar components along the major and the minor axis
respectively.\\  
{\em iii}) The spatial distribution of the old population
is, in contrast with the intermediate-age one, asymmetric along the two
axes. According to these evidence it seems that the relatively young
stellar population is more centrally concentrated than the old one.
This finding confirms preliminary empirical evidence for Carina suggested 
by Mighell (1997) and, more strongly, by Harbeck et al. (2001) and 
further supports the 
occurrence of different radial distributions in dSphs that present 
multiple stellar components (Stetson et al. 1998; Harbeck et al. 2001).
Moreover, the old population presents a steeper gradient along the
South-West than the North-East direction. A well-defined "shoulder"
is present approximately 4-6 arcmin along the North-East direction
both along the major and the minor axis. According to recent numerical
experiments (Johnston, Sigurdsson, \& Hernquist 1999) a robust signature
of the tidal interaction of a dwarf satellite with the Galaxy is the
occurrence of a break in the slope of star count profile at the location
where unbound stars dominate (see their Figs. 15 and 16). If the shoulder
we detected is the empirical indicator of the predicted break then
current finding supplies a qualitative support to the large star
mass-loss rate, roughly 30\% Gyr$^{-1}$, that Johnston et al. (1999)
predict for Carina.

More detailed information
concerning the Carina orbit and its velocity are required to figure
out whether dSph companions are dominated by dark matter halos or by
the tidal interaction with the Galaxy (Kuhn \& Miller 1989;
Klessen \& Zhao 2002). The empirical scenario concerning Carina
is still puzzling, since both IH and Majewski et al. (2000) found 
evidence of an extra tidal component at a radius of $\approx 20$ arcmin. 
This supports the evidence that Carina is interacting with the tidal 
field of the Galaxy. On the other hand, Walcher et al. (2002) in a recent 
investigation based on CCD observations that cover a wide area (4 square 
degrees) around Carina did not detect the break in the slope of the density
profile. This suggests that Carina might be dominated by a dark matter halo, 
whereas our finding tends to support the first working hypothesis. 
Current empirical scenario does not help to settle this fundamental 
problem, and indeed detailed star counts well behind the tidal radius 
support the evidence that Ursa Minor dSph is undergoing a strong tidal 
interaction with the Milk Way (Kleyna et al. 1998; 
Martinez-Delgado et al. 2001). On the other hand, similar investigations 
on the Draco dSph do not find evidence of tidally induced tails (Aparicio, 
Carrera, \& Martinez-Delgado 2001; Odenkirchen et al. 2001; 
Piatek et al. 2002). 

However, current data do cover only a limited region around the center 
of the system. It is worth mentioning that we have already collected 
B and V time series data of the same field with the prime focus Mosaic 
Imager at CTIO 4m Blanco telescope. These data will be presented in a 
forthcoming paper together with multiband photometric data available 
in the ESO archive. This is just the beginning of the story, new deeper 
and accurate photometric databases already secured by our group will 
certainly supply more tight constraints on still open problems concerning
the stellar population content of the Carina galaxy.
It goes without saying that the difference in both the chemical
composition and in the radial distribution among old, intermediate-age,
and relatively young populations will certainly benefit of new and
massive spectroscopic measurements.
An observational challenge that the new multifiber spectrographs
available in telescopes of the 4m (Hydra) and 8m (Flames/Giraffe)
class should allow us to cope with in the near future.

\begin{acknowledgements}
It is a pleasure to thank Frank Valdes for many helpful suggestions
and advice on installing the {\it mscred} package we used for the
pre-processing of the WFI images. We acknowledge W. Gieren and 
G. Pietrzynski as well as the 2p2 ESO team and in particular R. Mendez 
and F. Selman for insightful discussions concerning the treatment of
WFI data. We wish also to thank an anonymous referee for his/her
detailed suggestions and helpful comments that improved both the
content and the readability of this paper. We are grateful to 
S. Cassisi, M. Cignoni, S. Degl'Innocenti, and P. G. Prada Moroni 
for sending us detailed sets of isochrones and HB models as well 
as to D. Hoard for carrying out the Jan 31 2000 observations. 
This research has made use of NASA's Astrophysics Data System Abstract
Service and of SIMBAD database operated at CDS, Strasbourg, France. 
This work was supported by MIUR/Cofin~2000 under the project: 
"Stellar Observables of Cosmological Relevance". HAS thanks the US 
National Science Foundation for support under the grant AST99-86943. 
\end{acknowledgements}

\pagebreak

\footnotesize{
\include{monelli.tables}
}

\begin{figure}
\caption{Carina true color image obtained using the {\em iraf} task 
{\em color.rgbsun}. The image covers a field of approximately 1 degree$^2$ 
centered on Carina and it was created by combining the deep B,V median images 
of the central regions ($\approx 0.3$ degree$^2$) collected with the 
WFI and the MOSAIC Imager at the 4m Blanco/CTIO 
telescopes, together with the shallower B,V,I median images collected by 
EIS with the WFI. Northern and Eastern directions are on top and on the left, 
respectively. (Only available at "Carina Project" website)\label{fig1}}
\end{figure}

\begin{figure}
\caption{Map in Galactic coordinates of the $68,000$ stars identified
in the central Carina field. The gaps mark the individual chips and are
present becaused B and V images were not dithered. Empty circles are due
to saturated Galactic field stars. The cross marks the Carina center
(see Table 1), while the orientation is showed in the bottom right
corner.\label{fig2}}
\end{figure}

\begin{figure}
\caption{Difference in right ascension and declination between current 
astrometric positions, obtained using the GSC2 catalog, and the UCAC1 
catalog. Note that the bulk of the 107 stars in our sample presents a 
discrepancy smaller, at 1$\sigma$ level, than 100 {\em mas}.\label{fig3}}
\end{figure}

\begin{figure}
\caption{Relative response of the 8 chips of the WFI mosaic.
The difference in instrumental V magnitude of the same stars as a function
of V magnitude measured in each chip and in the reference chip \# 56.
The unexpected large scatter in the residuals for the top chips \#50-\#53
suggests a dependence of the residuals on the position of the stars in
the mosaic.\label{fig4}}
\end{figure}

\begin{figure}
\caption{Relative response of the 8 chips of the WFI mosaic. The standard
stars from Stetson (120-140) in the Rubin 149 field measured in the 8 chips
in the V band have been corrected following the procedure described in \S 4
 and calibrated. The residuals now present a homogeneous pattern independent
of the position inside the chips and the scatter attains values
connected with the S/N ratio of the individual measurements.\label{fig5}}
\end{figure}

\begin{figure}
\caption{Same as Fig. 5, but as a function of the (B-V) color.\label{fig6}}
\end{figure}

\begin{figure}
\caption{Same as Fig. 5, but for the B magnitude.\label{fig7}}
\end{figure}

\begin{figure}
\caption{Same as Fig. 6, but for the B magnitude.\label{fig8}}
\end{figure}

\begin{figure}
\caption{Comparison between the (B-V) colors (top) and the V magnitudes
(bottom) of Stetson's standard stars brighter than V=17 mag and current
photometry and calibrations.\label{fig9}}
\end{figure}

\begin{figure}
\caption{Comparison between two independent absolute calibrations. A set
of isolated and good S/N stars have been selected as secondary standards.
These stars have been measured and calibrated independently on the B,V data
collected in December 1999 and on the B,V data collected in March 2000. The
residuals show a good agreement both in magnitude and in color.\label{fig10}}
\end{figure}

\begin{figure}
\caption{Comparison between (B-V) colors (top) and V magnitudes (bottom) of
current secondary standards and the secondary standards collected with the
0.9 m CTIO telescope. The two independent calibrations are consistent within
the errors.\label{fig11}}
\end{figure}

\begin{figure}
\caption{Comparison  between the (B-V) colors (top) and the V magnitudes
(bottom) by Hurley-Keller et al. (1998) and current photometry.\label{fig12}}
\end{figure}

\begin{figure}
\caption{The CMD of Carina central regions across the eight individual
chips of the WFI. The orientation is the same as in Fig. 2.\label{fig13}}
\end{figure}

\begin{figure}
\caption{The cumulative CMD of Carina central area. The parameters adopted to 
select the bona fide stars are $\mid sharpness \mid \le 1$ and $\chi \le 3$.
\label{fig14}}
\end{figure}

\begin{figure}
\caption{The same as Fig. 14, but together with Carina stars (dots) are 
also plotted the stars belonging to the old LMC cluster Reticulum 
(red open circles, Monelli et al. 2002). The latter sample was artificially 
shifted in magnitude by $\approx 1.68$ mag to match the old Carina 
population.\label{fig15}}
\end{figure}

\begin{figure}
\vspace*{1.5truecm}
\caption{Top - Comparison between theoretical predictions and empirical 
data for the GGC NGC~1904. The solid line shows an isochrone of 11 Gyr, 
constructed by adopting Y=0.23, Z=0.004, and a mixing length $\alpha=2$. 
The dashed line displays the ZAHB for the same chemical composition 
(see text for details). The photometry was performed by Piotto et al. 
(2002, see also the web page http://www.menhir.pd.astro.it). 
Bottom - Same as the top, but the data refer to Reticulum 
(Monelli et al. 2002).  Theoretical models were plotted using 
distance moduli and reddening estimates available in the literature. 
See text for more details.\label{fig16}}  
\end{figure}

\begin{figure}
\caption{The cumulative CMD of Carina together with theoretical isochrones
(solid blue lines) at fixed chemical composition -Y=0.23, Z=0.0004- 
and mixing length $\alpha =2$. The isochrones range from 0.6 to 11 
Gyr (see labeled values). Dashed line shows the ZAHB for the same chemical 
composition and for a progenitor age of 12 Gy. Note that the isochrone for 
$t\approx5$ Gyr nicely fits the red clump region. Red open circles display 
the Reticulum ridge line. See text for details concerning distance and 
reddening correction.\label{fig17}}
\end{figure}

\begin{figure}
\vspace*{1.5truecm}
\caption{Comparison between predicted and empirical He-burning stars.
Together with static stars (small dots) are also plotted the Carina 
variables: circles RR Lyrae stars, triangles, ACs. Crosses mark 
variables that present poor-phase coverage. Different line styles 
display predicted Zero Age He-burning structures at fixed chemical 
composition and progenitor ages ranging from 12 ($M=0.8 M_\odot$) 
to 0.6 ($M=2.2 M_\odot$) Gyr.\label{fig18}}
\end{figure}

\clearpage
\begin{figure}
\caption{The dashed and the solid boxes show the CMD regions selected as
representative of the old (HB and SGB) and of the young (RC, intermediate-age 
MS) population respectively. Variables have been plotted using the same symbols
as in Fig. 18.\label{fig19}}
\end{figure}

\begin{figure}
\caption{The isodensity map of the Carina stellar content (dashed contours).
The solid contours show the isodensity levels of the old (HB and SGB)
stellar population. The isodensity levels range from 20\% to 100\% with
a step of 20\%.\label{fig20}}
\end{figure}

\begin{figure}
\caption{Same as Fig. 20, but the solid contours show the isodensity levels
of the young (RC and intermediate-age MS) stellar population.\label{fig21}}
\end{figure}

\begin{figure}
\vspace*{1.5truecm}
\caption{Histograms of the star counts along the major (top) and the minor
(bottom) axis for the entire stellar population of the Carina dSph. 
The sigma of the gaussian fits are labeled.\label{fig22}}
\end{figure}

\begin{figure}
\vspace*{1.5truecm}
\caption{Histograms of the star counts for the old (top) and the 
intermediate-age (bottom) stellar populations (see text for details) 
along the Carina major axis.\label{fig23}}
\end{figure}

\begin{figure}
\vspace*{1.5truecm}
\caption{Same as Fig. 23, but along the Carina minor axis. As in the
case of the major axis, the old population shows a broad distribution.
\label{fig24}}
\end{figure}

\begin{figure}
\vspace*{1.5truecm}
\caption{Same as Fig. 22, but for blue plume stars along the major (top) 
and the minor (bottom) axis.  
\label{fig25}}
\end{figure}

\end{document}

%% file: monelli.tables.tex
\begin{center}
 
\tablewidth{0pt}
\begin{deluxetable}{lrc}
\tablecaption{Positional, photometric and structural parameters of the 
Carina dSph.}\label{tbl-1}
\tablehead{
\colhead{Parameter}&
\colhead{         }&   
\colhead{Ref.\tablenotemark{a}} }
\startdata
$\alpha$ (J2000)                          &   06~41~37    &     1    \nl  
$\delta$ (J2000)                          &  -50~58~00    &     1    \nl  
$M_V$ (mag)\tablenotemark{b}              &    -8.9       &     2    \nl  
$r_c$ (arcmin)\tablenotemark{c}           &   $11.96\pm1.5/14$\tablenotemark{d} &     2    \nl  
$r_t$ (arcmin)\tablenotemark{e}           &  $22.54\pm1.4/32$\tablenotemark{d} &     2    \nl  
$e$\tablenotemark{f}                      &  $0.32\pm0.04$&     2    \nl  
$PA$ (deg)\tablenotemark{g}           &  $64\pm2.5$       &     2    \nl  
$\sigma_V$ (km~s$^{-1}$)\tablenotemark{h} &  $6.8\pm1.6$  &     3    \nl  
[Fe/H]\tablenotemark{i}                   &$-2.0\pm0.30$  &     4    \nl  
E(B-V)\tablenotemark{j}                   &$0.04\pm0.02$  &     1    \nl  
$(m-M)_0$ (mag)\tablenotemark{k}          &$20.03\pm0.09$ &     1    \nl  
\enddata 
\tablenotetext{a}{References: 1) Mateo 1998; 2) Walcher et al. 2002;
3) Mateo et al. 1993; 4) Smecker-Hane et al. 1999a. 
\hspace*{0.5mm}$^b$ Total Visual magnitude.  
\hspace*{0.5mm}$^c$ Core radius. 
\hspace*{0.5mm}$^d$ The two estimates refer to an empirical
(King 1962) and to a theorethical (Binney \& Tremaine 1987) profile.
\hspace*{0.5mm}$^e$ Tidal radius.  
\hspace*{0.5mm}$^f$ Eccentricity. 
\hspace*{0.5mm}$^g$ Major axis position angle. 
\hspace*{0.5mm}$^h$ Stellar central velocity, dispersion.  
\hspace*{0.5mm}$^i$ Metallicity. 
\hspace*{0.5mm}$^j$ Reddening.  
\hspace*{0.5mm}$^k$ True distance modulus.}   
\end{deluxetable}

\tablewidth{0pt}
\begin{deluxetable}{llrccc}
\tablecaption{Photometric data adopted in this investigation.}\label{tbl-1}
\tablehead{
\colhead{Field}&
\colhead{Instrument}&   
\colhead{Images}&  
\colhead{FOV\tablenotemark{a}} &  
\colhead{R.A.\tablenotemark{b}}&  
\colhead{DEC.\tablenotemark{b}} 
}
\startdata
C     & WFI@2.2m\tablenotemark{c} & 54(B,V)  & $34'\times33'$ & 06~41~37 &-50~58~00 \\
C     & MC@4m\tablenotemark{d}   & 50(B,V)  & $36'\times36'$ & \ldots& \ldots\\
EIS-a & WFI@2.2m\tablenotemark{e} & 2(B,V,I) & $34'\times33'$ & 06~42~36 & -50~45~00\\
EIS-b & WFI@2.2m\tablenotemark{e} & 2(B,V,I) & $34'\times33'$ & 06~38~55 & -50~45~00\\
EIS-c & WFI@2.2m\tablenotemark{e} & 2(B,V,I) & $34'\times33'$ & 06~38~55 & -51~11~00\\
EIS-d & WFI@2.2m\tablenotemark{e} & 2(B,V,I) & $34'\times33'$ & 06~42~36 & -51~11~00\\
\enddata 
\tablenotetext{a}{Field of view (arcmin). 
\hspace*{0.5mm}$^b$Field coordinates (J2000). 
\hspace*{0.5mm}$^c$ Data set collected with the WFI available at the 2.2m 
ESO/MPI telescope. The photometry is discussed in this paper.
\hspace*{0.5mm}$^d$ Data set collected with the Mosaic Camera available 
at the 4m CTIO Blanco telescope. These data were adopted to produce Fig. 1, 
the photometry will be discussed in a forthcoming paper.
\hspace*{0.5mm}$^e$ Data set collected with the WFI by the EIS project. 
telescope. These data were adopted to produce Fig. 1 and to improve the 
accuracy of the astrometric solution (see \S 2.3). The photometry will 
be discussed in a forthcoming paper.}   
\end{deluxetable}

\clearpage
\tablewidth{0pt}
\begin{deluxetable}{ccccccc}
\tablecaption{Log of scientific CCD images of Carina}\label{tbl-1}
\tablehead{
\colhead{Frame}&
\colhead{Date\tablenotemark{a}}&
\colhead{JD\tablenotemark{b}}&
\colhead{UT Start\tablenotemark{c}}&
\colhead{Exposure\tablenotemark{d}}&
\colhead{Filter\tablenotemark{e}}&
\colhead{Seeing\tablenotemark{f}} \nl
\colhead{(1)}&
\colhead{(2)}&
\colhead{(3)}&
\colhead{(4)}&
\colhead{(5)}&
\colhead{(6)}&
\colhead{(7)} } 
\startdata
           \multicolumn{7}{c}{Central field\tablenotemark{g}} \nl  
46253 & Jan5 & 48.0554 & 1:19:51 & 300 & V & 1.7 \nl
46254 & Jan5 & 48.0607 & 1:27:29 & 300 & B & 1.5 \nl
46358 & Jan5 & 49.0162 & 0:23:25 &  30 & V & 1.4 \nl
46359 & Jan5 & 49.0211 & 0:30:27 &  30 & V & 1.5 \nl
46364 & Jan5 & 49.0648 & 1:33:19 & 500 & V & 1.7 \nl
46365 & Jan5 & 49.0724 & 1:44:18 & 500 & B & 1.2 \nl
46366 & Jan5 & 49.0792 & 1:54:07 & 500 & V & 1.1 \nl
46367 & Jan5 & 49.0872 & 2:05:39 & 500 & B & 1.1 \nl
46368 & Jan5 & 49.0949 & 2:16:43 & 500 & V & 1.0 \nl
46369 & Jan5 & 49.1033 & 2:28:50 & 500 & B & 1.1 \nl
46370 & Jan5 & 49.1120 & 2:41:21 & 500 & V & 1.1 \nl
46371 & Jan5 & 49.1206 & 2:52:21 & 500 & B & 1.2 \nl
46372 & Jan5 & 49.1268 & 3:02:40 & 500 & V & 1.1 \nl
46373 & Jan5 & 49.1345 & 3:13:48 & 500 & B & 1.2 \nl
46374 & Jan5 & 49.1418 & 3:24:12 & 500 & V & 1.0 \nl
46375 & Jan5 & 49.1486 & 3:34:07 & 500 & B & 1.1 \nl
46376 & Jan5 & 49.1556 & 3:44:08 & 500 & V & 0.9 \nl
46377 & Jan5 & 49.1624 & 3:53:57 & 500 & B & 0.9 \nl
46378 & Jan5 & 49.1693 & 4:03:51 & 500 & V & 0.8 \nl
46379 & Jan5 & 49.1761 & 4:13:36 & 500 & B & 0.8 \nl
46382 & Jan5 & 49.1969 & 4:43:38 & 500 & B & 0.9 \nl
46383 & Jan5 & 49.2040 & 4:53:49 & 500 & V & 0.9 \nl
46384 & Jan5 & 49.2108 & 5:03:34 & 500 & B & 0.9 \nl
46385 & Jan5 & 49.2181 & 5:14:11 & 500 & V & 0.8 \nl
46386 & Jan5 & 49.2285 & 5:29:02 & 500 & B & 0.9 \nl
46387 & Jan5 & 49.2363 & 5:40:20 & 500 & V & 0.9 \nl
46388 & Jan5 & 49.2431 & 5:50:11 & 500 & B & 0.8 \nl
46391 & Jan5 & 49.2593 & 6:13:27 & 500 & V & 0.8 \nl
46392 & Jan5 & 49.2662 & 6:23:24 & 500 & B & 0.9 \nl
46393 & Jan5 & 49.2731 & 6:33:18 & 500 & V & 0.8 \nl
46394 & Jan5 & 49.2799 & 6:43:06 & 500 & B & 0.9 \nl
46395 & Jan5 & 49.2873 & 6:53:50 & 500 & V & 0.9 \nl
46396 & Jan5 & 49.2942 & 7:03:40 & 500 & B & 1.0 \nl
46397 & Jan5 & 49.3011 & 7:13:43 & 500 & V & 0.9 \nl
46398 & Jan5 & 49.3079 & 7:23:28 & 500 & B & 1.0 \nl
46399 & Jan5 & 49.3148 & 7:33:23 & 500 & V & 0.9 \nl
46400 & Jan5 & 49.3216 & 7:43:12 & 500 & B & 1.1 \nl
46401 & Jan5 & 49.3287 & 7:53:28 & 500 & V & 1.0 \nl
46402 & Jan5 & 49.3357 & 8:03:33 & 500 & B & 1.2 \nl
46403 & Jan5 & 49.3425 & 8:13:15 & 500 & V & 1.0 \nl
46404 & Jan5 & 49.3494 & 8:23:13 & 500 & B & 1.3 \nl
46405 & Jan5 & 49.3564 & 8:33:19 & 500 & V & 1.0 \nl
46406 & Jan5 & 49.3632 & 8:43:07 & 500 & B & 1.3 \nl
46509 & Jan6 & 50.0440 & 1:03:26 & 500 & V & 1.4 \nl
46510 & Jan6 & 50.0525 & 1:15:41 & 500 & B & 1.4 \nl
46511 & Jan6 & 50.0598 & 1:26:09 & 900 & V & 1.5 \nl
46512 & Jan6 & 50.0727 & 1:44:46 & 900 & B & 1.4 \nl
46513 & Jan6 & 50.0856 & 2:03:16 & 900 & V & 1.3 \nl
46514 & Jan6 & 50.0978 & 2:20:53 & 900 & B & 1.3 \nl
46516 & Jan6 & 50.1180 & 2:50:03 & 600 & V & 1.1 \nl
46517 & Jan6 & 50.1271 & 3:03:06 & 600 & B & 1.2 \nl
46518 & Jan6 & 50.1361 & 3:16:04 & 600 & V & 1.1 \nl
46520 & Jan6 & 50.1548 & 3:43:02 & 500 & B & 1.1 \nl
46521 & Jan6 & 50.1630 & 3:54:44 & 500 & V & 1.1 \nl
46522 & Jan6 & 50.1717 & 4:07:23 & 500 & B & 1.2 \nl
46523 & Jan6 & 50.1790 & 4:17:50 & 500 & V & 0.9 \nl
46524 & Jan6 & 50.1860 & 4:27:53 & 500 & B & 0.9 \nl
46525 & Jan6 & 50.1933 & 4:38:27 & 500 & V & 0.8 \nl
46526 & Jan6 & 50.2004 & 4:48:37 & 500 & B & 0.8 \nl
46527 & Jan6 & 50.2074 & 4:58:47 & 500 & V & 0.9 \nl
46528 & Jan6 & 50.2156 & 5:10:33 & 500 & B & 1.0 \nl
46529 & Jan6 & 50.2228 & 5:20:51 & 144 & V & 1.0 \nl
46531 & Jan6 & 50.2345 & 5:37:42 & 500 & V & 0.8 \nl
46532 & Jan6 & 50.2427 & 5:49:36 & 500 & B & 0.8 \nl
46533 & Jan6 & 50.2500 & 5:59:56 & 500 & V & 0.8 \nl
46534 & Jan6 & 50.2584 & 6:12:11 & 500 & B & 0.8 \nl
46535 & Jan6 & 50.2654 & 6:22:11 & 500 & V & 0.8 \nl
46536 & Jan6 & 50.2727 & 6:32:48 & 500 & B & 0.9 \nl
46537 & Jan6 & 50.2796 & 6:42:43 & 500 & V & 0.8 \nl
46538 & Jan6 & 50.2867 & 6:52:57 & 500 & B & 1.0 \nl
46540 & Jan6 & 50.3007 & 7:13:04 & 156 & V & 1.0 \nl
46541 & Jan6 & 50.3048 & 7:18:55 & 500 & V & 0.9 \nl
46544 & Jan6 & 50.3279 & 7:52:13 & 500 & V & 1.1 \nl
46545 & Jan6 & 50.3361 & 8:04:00 & 600 & B & 1.3 \nl
46546 & Jan6 & 50.3440 & 8:15:23 & 600 & V & 1.0 \nl
46547 & Jan6 & 50.3521 & 8:27:05 & 700 & B & 1.2 \nl
46548 & Jan6 & 50.3623 & 8:41:49 & 120 & B & 1.4 \nl
46549 & Jan6 & 50.3649 & 8:45:28 &  30 & B & 1.2 \nl
46550 & Jan6 & 50.3673 & 8:48:58 & 120 & V & 1.1 \nl
46551 & Jan6 & 50.3694 & 8:52:04 &  30 & V & 1.1 \nl
46602 & Jan7 & 51.0407 & 0:58:44 & 500 & V & 0.9 \nl
46603 & Jan7 & 51.0477 & 1:08:43 & 500 & B & 0.9 \nl
46604 & Jan7 & 51.0548 & 1:18:55 & 500 & V & 1.0 \nl
46605 & Jan7 & 51.0620 & 1:29:22 & 500 & B & 1.0 \nl
46606 & Jan7 & 51.0688 & 1:39:05 & 500 & V & 0.9 \nl
46607 & Jan7 & 51.0755 & 1:48:45 & 500 & B & 0.9 \nl
46608 & Jan7 & 51.0822 & 1:58:25 & 500 & V & 0.9 \nl
46609 & Jan7 & 51.0890 & 2:08:15 & 500 & B & 1.0 \nl
46610 & Jan7 & 51.0981 & 2:21:17 & 500 & V & 0.9 \nl
46611 & Jan7 & 51.1049 & 2:31:06 & 500 & B & 1.0 \nl
46614 & Jan7 & 51.1229 & 2:57:07 & 500 & V & 0.7 \nl
46615 & Jan7 & 51.1297 & 3:06:49 & 500 & B & 0.8 \nl
46616 & Jan7 & 51.1364 & 3:16:32 & 500 & V & 0.9 \nl
46617 & Jan7 & 51.1432 & 3:26:17 & 500 & B & 0.8 \nl
46618 & Jan7 & 51.1500 & 3:36:04 & 500 & V & 0.8 \nl
46619 & Jan7 & 51.1568 & 3:45:47 & 500 & B & 0.9 \nl
46620 & Jan7 & 51.1635 & 3:55:34 & 500 & V & 0.8 \nl
46621 & Jan7 & 51.1703 & 4:05:18 & 500 & B & 0.8 \nl
46622 & Jan7 & 51.1770 & 4:14:58 & 500 & V & 0.8 \nl
46623 & Jan7 & 51.1838 & 4:24:42 & 500 & B & 0.9 \nl
46624 & Jan7 & 51.1906 & 4:34:32 & 500 & V & 1.0 \nl
46625 & Jan7 & 51.1974 & 4:44:22 & 500 & B & 1.0 \nl
46626 & Jan7 & 51.2042 & 4:54:04 & 500 & V & 0.8 \nl
46627 & Jan7 & 51.2110 & 5:03:54 & 500 & B & 1.0 \nl
46628 & Jan7 & 51.2202 & 5:17:12 & 500 & V & 1.0 \nl
46629 & Jan7 & 51.2270 & 5:26:56 & 500 & B & 1.3 \nl
46630 & Jan7 & 51.2337 & 5:36:34 & 500 & V & 1.2 \nl
46631 & Jan7 & 51.2405 & 5:46:21 & 500 & B & 1.3 \nl
46632 & Jan7 & 51.2472 & 5:56:05 & 500 & V & 1.0 \nl
46633 & Jan7 & 51.2539 & 6:05:44 & 500 & B & 1.2 \nl
46634 & Jan7 & 51.2607 & 6:15:28 & 500 & V & 1.1 \nl
46635 & Jan7 & 51.2675 & 6:25:14 & 500 & B & 1.2 \nl
46637 & Jan7 & 51.2808 & 6:44:29 & 500 & V & 0.8 \nl
46638 & Jan7 & 51.2876 & 6:54:12 & 500 & B & 0.9 \nl
46639 & Jan7 & 51.2944 & 7:03:58 & 500 & V & 0.8 \nl
46640 & Jan7 & 51.3015 & 7:14:13 & 500 & B & 1.0 \nl
46641 & Jan7 & 51.3084 & 7:24:14 & 500 & V & 0.9 \nl
46642 & Jan7 & 51.3162 & 7:35:21 & 500 & B & 1.1 \nl
46643 & Jan7 & 51.3232 & 7:45:28 & 500 & V & 1.2 \nl
46644 & Jan7 & 51.3303 & 7:55:39 & 500 & B & 1.0 \nl
46645 & Jan7 & 51.3372 & 8:05:33 & 900 & V & 0.9 \nl
46646 & Jan7 & 51.3492 & 8:22:54 & 900 & B & 1.1 \nl
46647 & Jan7 & 51.3616 & 8:40:46 & 500 & B & 1.1 \nl
           \multicolumn{7}{c}{EIS-a field\tablenotemark{g}} \nl  
52197 & Feb26&  100.042209  &  1:00:47  & 240 & B & 1.6  \nl 
52198 & Feb26&  100.046436  &  1:06:52  & 240 & B & 1.5  \nl 
52201 & Feb26&  100.053631  &  1:17:14  & 240 & V & 1.4  \nl 
52202 & Feb26&  100.057854  &  1:23:19  & 240 & V & 1.6  \nl 
52204 & Feb26&  100.063370  &  1:31:15  & 240 & I & 1.5  \nl 
52205 & Feb26&  100.067260  &  1:36:51  & 240 & I & 1.5  \nl 
          \multicolumn{7}{c}{EIS-b field\tablenotemark{g}} \nl  
52213 & Feb26&  100.096887  &  2:17:56  &  30 & B & 1.9  \nl 
52214 & Feb26&  100.101842  &  2:19:31  & 240 & B & 1.8  \nl 
52210 & Feb26&  100.084558  &  2:01:46  & 240 & V & 1.7  \nl 
52211 & Feb26&  100.090131  &  2:09:47  & 240 & V & 1.8  \nl 
52207 & Feb26&  100.073152  &  1:45:20  & 240 & I & 1.7  \nl 
52208 & Feb26&  100.076994  &  1:50:52  & 240 & I & 1.7  \nl 
           \multicolumn{7}{c}{EIS-c field\tablenotemark{g}} \nl  
52217 & Feb26&  100.107976  &  2:35:29  & 240 & B & 2.0  \nl 
52218 & Feb26&  100.112669  &  2:42:15  & 240 & B & 1.9  \nl 
52220 & Feb26&  100.119605  &  2:52:14  & 240 & V & 1.9  \nl 
52221 & Feb26&  100.123418  &  2:57:43  & 240 & V & 1.9  \nl 
52225 & Feb26&  100.137814  &  3:18:27  & 240 & I & 0.9  \nl 
52226 & Feb26&  100.141563  &  3:23:51  & 240 & I & 0.9  \nl 
           \multicolumn{7}{c}{EIS-d field\tablenotemark{g}} \nl  
52234 & Feb26&  100.167139  &  4:00:41  & 240 & B & 1.7  \nl 
52235 & Feb26&  100.170812  &  4:05:58  & 240 & B & 1.7  \nl 
52231 & Feb26&  100.157976  &  3:47:29  & 240 & V & 1.7  \nl 
52232 & Feb26&  100.161693  &  3:52:50  & 240 & V & 1.4  \nl 
52228 & Feb26&  100.148599  &  3:33:59  & 240 & I & 1.5  \nl 
52229 & Feb26&  100.152297  &  3:39:19  & 240 & I & 1.5  \nl 
\enddata 
\tablenotetext{a}{UT Date 2000.
\hspace*{0.5mm}$^b$ Julian Day - 2451500.0 
\hspace*{0.5mm}$^c$ UT start of the exposure time.
\hspace*{0.5mm}$^d$ Individual exposure time (sec).
\hspace*{0.5mm}$^e$ The B and the V filters are the B/99 and the V/89 
ESO filters. 
\hspace*{0.5mm}$^f$ Individual seeing (arcsec).
\hspace*{0.5mm}$^g$ See Table 2 for more details.} 
\end{deluxetable}

\tablewidth{0pt}
\begin{deluxetable}{llrc}
\tablecaption{Standard stellar fields observed for this investigation.}\label{tbl-1}
\tablehead{
\colhead{Set}&
\colhead{Field}&
\colhead{Instrument}&   
\colhead{Images} 
}
\startdata
 Dec\tablenotemark{a}& Rubin~149 & WFI@2.2m & (B,V) \\
 Jan\tablenotemark{b}& SA~98, Rubin~149, PG~1047& 0.9m CTIO  & (B,V) \\
 Mar\tablenotemark{c}& Rubin~149, PG~0918, PG~1323, PG~1633& WFI@2.2m & (B,V)\\
\enddata 
\tablenotetext{a}{Data set collected between Dec 1999 and Jan 2000 with the 
WFI at the 2.2m ESO/MPI telescope. 
\hspace*{0.5mm}$^b$ Data sets collected on Jan 31 at the 0.9m CTIO telescope. 
\hspace*{0.5mm}$^c$ Data sets collected on March 2000 with the WFI.}   
\end{deluxetable}

\tablewidth{0pt}
\begin{deluxetable}{rlrcl}
\tablecaption{Log of standard stellar fields and scientific CCD images.}\label{tbl-1}
\tablehead{
\colhead{Frame}&
\colhead{Date\tablenotemark{a}}&
\colhead{Exposure\tablenotemark{b}}&
\colhead{Filter}&
\colhead{Field\tablenotemark{c}} \nl
\colhead{(1)}&
\colhead{(2)}&
\colhead{(3)}&
\colhead{(4)}&
\colhead{(5)} } 
\startdata
           \multicolumn{5}{c}{December set\tablenotemark{d}} \nl  
45642 & Dec31&  30 & V & Rubin~149  \nl
45649 & Dec31&  30 & V & Rubin~149  \nl
45662 & Dec31&  30 & V & Rubin~149  \nl
45650 & Dec31&  30 & V & Rubin~149  \nl
45665 & Dec31&  30 & V & Rubin~149  \nl
45641 & Dec31&  30 & V & Rubin~149  \nl
45659 & Dec31&  30 & V & Rubin~149 \nl
45648 & Dec31&  30 & V & Rubin~149  \nl
45668 & Dec31&  30 & V & Rubin~149  \nl
45655 & Dec31&  30 & V & Rubin~149  \nl
45647 & Dec31&  20 & B & Rubin~149  \nl
45657 & Dec31&  20 & B & Rubin~149 \nl
45656 & Dec31&  20 & B & Rubin~149 \nl
45643 & Dec31&  30 & B & Rubin~149 \nl
45663 & Dec31&  20 & B & Rubin~149  \nl
45664 & Dec31&  20 & B & Rubin~149  \nl
45651 & Dec31&  20 & B & Rubin~149  \nl
45644 & Dec31&  20 & B & Rubin~149  \nl
45658 & Dec31&  20 & B & Rubin~149  \nl
45670 & Dec31&  20 & B & Rubin~149  \nl
45669 & Dec31&  20 & B & Rubin~149  \nl
45788 & Jan1 & 300 & V & Carina OC  \nl
45789 & Jan1 & 300 & V & Carina OC  \nl
45790 & Jan1 & 300 & V & Carina OC  \nl
45791 & Jan1 & 300 & B & Carina OC  \nl
45792 & Jan1 & 300 & B & Carina OC  \nl
45793 & Jan1 & 300 & B & Carina OC  \nl

           \multicolumn{5}{c}{January set\tablenotemark{e}} \nl  
56   & Jan31 &  60 & B & SA~98 \nl 
57   & Jan31 &  60 & V & SA~98 \nl 
58   & Jan31 &  30 & B & SA~98 \nl 
59   & Jan31 &  30 & V & SA~98 \nl 
66   & Jan31 &  40 & B & Rubin~149 \nl
67   & Jan31 &  40 & B & Rubin~149 \nl
68   & Jan31 &  20 & V & Rubin~149 \nl
69   & Jan31 &  20 & V & Rubin~149 \nl
84   & Jan31 &  60 & B & PG~1047   \nl
85   & Jan31 &  60 & B & PG~1047   \nl
86   & Jan31 &  30 & V & PG~1047   \nl
87   & Jan31 &  30 & V & PG~1047   \nl 
60   & Jan31 & 400 & B & Carina C  \nl 
61   & Jan31 & 600 & V & Carina C  \nl 
62   & Jan31 & 400 & B & Carina C  \nl 
63   & Jan31 & 600 & V & Carina C  \nl 
64   & Jan31 & 400 & B & Carina C  \nl 
65   & Jan31 & 600 & V & Carina C  \nl 

           \multicolumn{5}{c}{March set\tablenotemark{d}} \nl  
53481 & Mar5 &   7 & B & Rubin~149  \nl
53482 & Mar5 &   7 & V & Rubin~149  \nl
53489 & Mar5 &  15 & B & PG~0918  \nl
53508 & Mar5 &  15 & B & PG~0918   \nl
53488 & Mar5 &   8 & V & PG~0918   \nl
53507 & Mar5 &  12 & V & PG~0918   \nl
53528 & Mar5 &   8 & B & PG~1323  \nl
53526 & Mar5 &   8 & V & PG~1323  \nl
53542 & Mar5 &  10 & B & PG~1633  \nl
53555 & Mar5 &  10 & B & PG~1633  \nl
53543 & Mar5 &  10 & V & PG~1633  \nl
53554 & Mar5 &  10 & V & PG~1633  \nl
53491 & Mar5 & 300 & B & Carina C \nl
53492 & Mar5 & 300 & V & Carina C  \nl
\enddata 
\tablenotetext{a}{UT Date for the December set 1999/2000, while 
for the January and the March set is 2000.
\hspace*{0.5mm}$^b$ Individual exposure time (sec).
\hspace*{0.5mm}$^c$ Observed field. The field "Carina C" is centered 
on the galaxy, while "Carina OC" is shifted by $\approx 25$ arcmin in 
right ascension, East direction.
\hspace*{0.5mm}$^d$ Data collected with the WFI at the 2.2m ESO/MPI 
telescope, field of view $34'\times33'$.  
\hspace*{0.5mm}$^e$ Data collected with the 0.9m CTIO telescope, 
field of view $14\times 14'$.} 
\end{deluxetable}
\end{center}